\documentclass[12]{article}
\usepackage{fleqn}
\usepackage{graphicx}
\begin{document}
\Large
\begin{center}
{\bf Mermin's Pentagram as an Ovoid of PG(3,\,2)}
\end{center}
\vspace*{.2cm}
\large
\begin{center}
Metod Saniga$^{1}$ and P\' eter L\' evay$^{2}$
\end{center}
\vspace*{-.5cm} \normalsize
\begin{center}
$^{1}$Astronomical Institute, Slovak Academy of Sciences\\
SK-05960 Tatransk\' a Lomnica\\ Slovak Republic\\
(msaniga@astro.sk)

\vspace*{.0cm} and

\vspace*{.0cm}
$^{2}$Department of Theoretical Physics, Institute of Physics\\
Budapest University of Technology and Economics\\
H-1521 Budapest, Hungary\\
(levay@neumann.phy.bme.hu)

\end{center}

\vspace*{.0cm} \noindent \hrulefill

\vspace*{.1cm} \noindent {\bf Abstract}

\noindent Mermin's pentagram, a specific set of ten three-qubit observables arranged in quadruples of pairwise commuting ones into five edges of a pentagram and used to provide a very simple proof of the Kochen-Specker theorem, is shown to be isomorphic to an ovoid (elliptic quadric) of the three-dimensional projective space of order two, PG$(3,2)$. This demonstration employs properties of the real three-qubit Pauli group embodied in the geometry of the symplectic polar space $W(5,2)$ and rests on the facts that: 1) the four observables/operators on any of the five edges of the pentagram can be viewed as points of an affine plane of order two, 2) all the ten observables lie on a hyperbolic quadric of the five-dimensional projective space of order two, PG$(5,2)$, and 3) that the points of this quadric are in a well-known bijective correspondence with the lines of PG$(3,2)$.

\vspace*{-.1cm} \noindent \hrulefill

\vspace*{.3cm}
\large

\section{Introduction}
In 1993, Mermin \cite{mer} proposed two ingenious and remarkably simple geometrical proofs of the Kochen-Specker theorem. The first one, which became known as Mermin's (magic) square, employs a set of nine elements/observables of the generalized Pauli group of {\it two}-qubits. These nine observables are placed at the vertices of a $3 \times 3$ grid and form six sets of three pairwise commuting elements, lying along three horizontal and three vertical lines, each observable thus pertaining to two such sets. The observables are selected in such a way that the product of their triples in five of the six sets is $+I$, whilst in the remaining set it is $-I$, $I$ being the identity matrix. The second proof, known as Mermin's (magic) pentagram, uses a set of ten elements of the {\it three}-qubit Pauli group. Here, the ten observables form five sets of four mutually commuting elements placed along the five edges of a pentagram. Again, each observable belongs to two such sets (``contexts") and the product of fours in any given sets is $+I$ except for one where it yields $-I$.

Soon after the properties of $N$-qubit Pauli groups were found to be fully encoded in the geometry of the symplectic polar spaces of rank $N$ and order two, $W(2N-1,2)$ \cite{sp,ps,hos},  the Mermin (magic) square could readily be ascribed a neat finite geometrical meaning, namely as: a special kind of geometric hyperplane of $W(3,2)$ [2,3], a hyperbolic quadric in PG$(3,2)$ [4], or a projective line over the direct product of two smallest Galois fields, $P_{1}(GF(2) \times GF(2))$ \cite{spp}. These last findings repeatedly turned out to be of great physical importance, for example, in revealing a fascinating finite-geometrical meaning of the $E_{6(6)}-$symmetric black hole entropy formula of $D=5$ supergravity theories \cite{lsvp}. In the present paper, employing basic properties of the finite symplectic polar space behind the three-qubit Pauli group, we shall show that the structure of the Mermin pentagrammatic configuration can similarly be recast in terms of another well-known object of finite geometry.

\begin{figure}[t]
\centerline{\includegraphics[width=7.3cm,clip=]{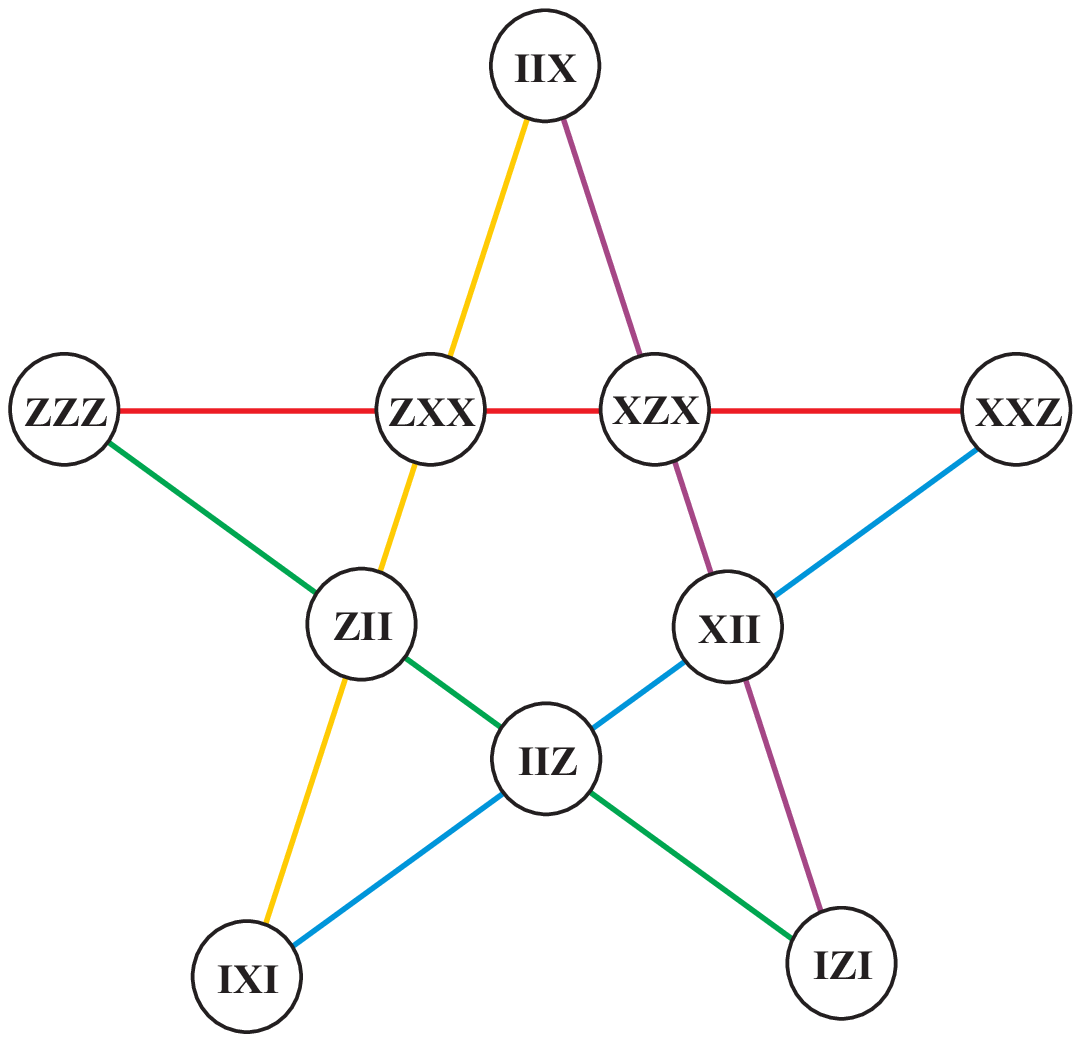}\includegraphics[width=7.cm,clip=]{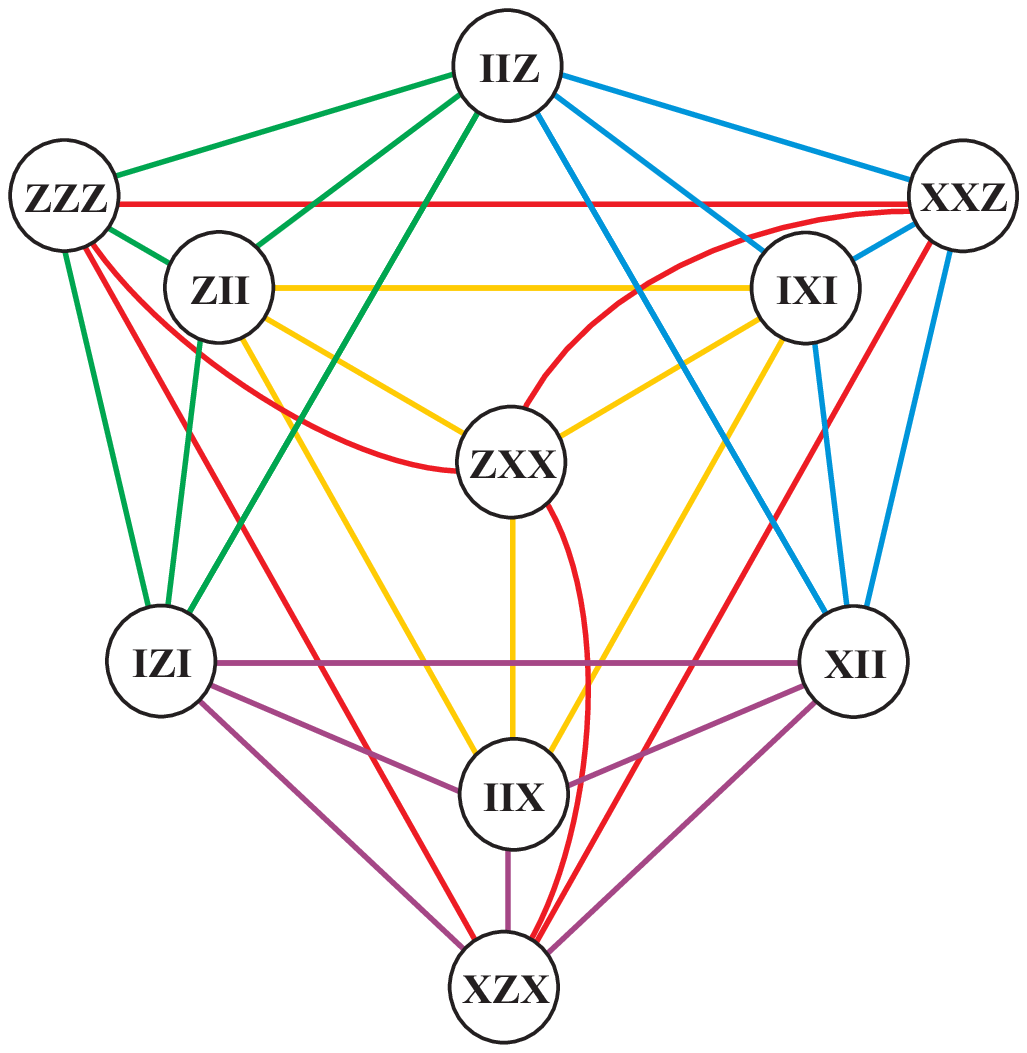}}

\vspace*{.2cm} \caption{{\it Left}: --- An illustration of the Mermin pentagram. The four three-qubit observables along any edge are pairwise commuting; the product of those along the horizontal (red) edge is $-I$, while those along any other edge multiply to $+I$. {\it Right}:
--- A picture of the finite geometric configuration behind the Mermin pentagram: the five edges of the pentagram correspond to five copies of the affine plane of order two, sharing pairwise a single point.}
\end{figure}

\section{Mermin's Pentagram, Affine Plane of Order Two, the Klein Quadric, the Klein Correspondence and an Ovoid of PG(3,\,2)}
Our starting point is a particular copy of the Mermin pentagram depicted in Figure 1, left; here $I$ is the $8 \times 8$ identity matrix,  $X \equiv \sigma_x, Z \equiv \sigma_z$, and, e.\,g., $ZZZ$ is a shorthand for $Z \otimes Z \otimes Z$.   The set of ten three-qubit operators we adopted is, however, not that of Mermin \cite{mer}, but that of Aravind \cite{ar}, who was motivated by the paper by Kernaghan and Peres \cite{kp}.  The reason is that all the ten matrices are real and can thus be viewed as a subset of the {\it real} three-qubit Pauli group.

\begin{figure}[t]
\centerline{\includegraphics[width=9.5cm,clip=]{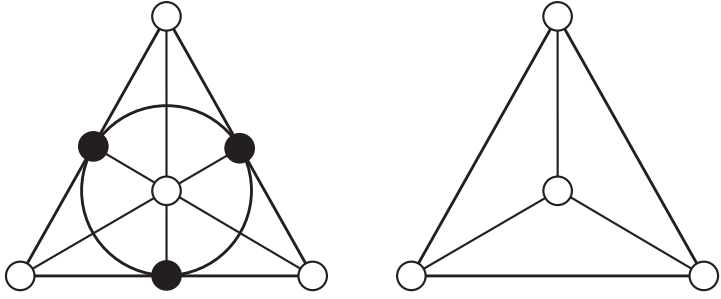}}
\vspace*{.2cm}
\caption{An illustration of the case (left) where no three of the four observables (represented by empty circles) pertaining to an edge  are collinear in the associated Fano plane; they form an affine plane of order two (right).}
\end{figure}
\begin{figure}[pth!]
\centerline{\includegraphics[width=9.5cm,clip=]{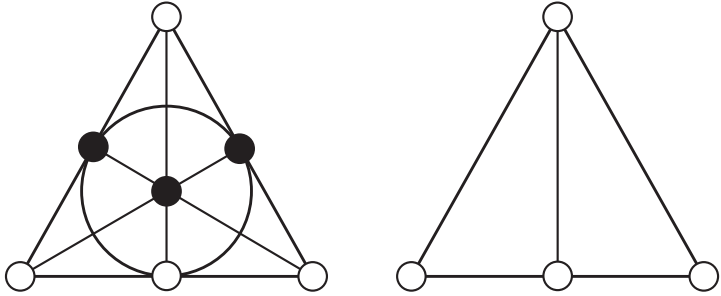}}
\vspace*{.2cm}
\caption{An illustration of a hypothetical case where three of the four observables would lie on a line of the Fano plane.}
\end{figure}

In order to ``decipher" the structure of our Mermin's pentagram one has to invoke certain aspects of the finite geometric interpretation of the real three qubit Pauli group \cite{sp,ps,hos}. Omitting the identity and disregarding signs, 63 elements of this group correspond to 63 points of the symplectic polar space $W(5,2)$, and maximal subsets of pairwise commuting elements of the group have their counterparts in maximal totally isotropic subspaces of $W(5,2)$, which are Fano planes. As each Fano plane has seven points, any maximal subset of mutually commuting operators is of cardinality seven. Each edge from our Mermin's pentagram, obviously, selects from its ambient Fano plane only four points. There are two possibilities for such a selection; either no three of these four points are collinear (Figure 2), or three of them lie on a line in the corresponding Fano plane (Figure 3). The second possibility can, however, be readily disregarded as the four observables on any edge have obviously the same footing. This implies that each edge of the pentagram is a copy of the affine plane of order two and the pentagram framework can be reformulated as five affine planes of order two sharing pairwise a single point, no three being on the same point --- see Figure 1, right. Next we employ the fact that we are dealing with the real three-qubit Pauli group. In such a case [4] the structure of the symplectic polar space $W(5,2)$ is refined in terms of the orthogonal polar space $Q^{+}(5,2)$ (which is nothing but the famous Klein quadric), the points of which correspond to the {\it symmetric} operators/elements of the group. As all ten observables of our pentagram are symmetric, they must lie on $Q^{+}(5,2)$. Now, the generators of $Q^{+}(5,2)$ (i.\,e., maximal subspaces fully lying on $Q^{+}(5,2)$) are Fano planes and there are two systems of them, each having 15 members. Any two distinct planes from the same system share a point, whilst two planes from different systems are either disjoint, or have a line in common (see, e.\,g., \cite{hir,con}). It then follows that our five affine planes must all originate from Fano planes of the same system.

As a final step, we employ the famous Klein correspondence between the points of the Klein quadric $Q^{+}(5,2)$ and the lines of PG$(3,2)$ (see, e.\,g.,  Table 15. 10 of \cite{hir} for more details). Under this correspondence, the seven points in a plane of one system of $Q^{+}(5,2)$ correspond to the seven lines through a point of PG$(3,2)$, and those of a plane of the other system to the seven lines lying in a plane of PG$(3,2)$; an affine plane of order two lying on $Q^{+}(5,2)$ will then have for its PG$(3,2)$-image either four lines through a point, no three coplanar, or four lines in a plane, no three concurrent.
Adopting the former view, we thus find that the PG$(5,2)$-configuration depicted in Figure 1, right, has for its PG$(3,2)$-counterpart the set of five points, no three collinear and no four lying in the same plane, that is, a copy of an {\it ovoid} (or, elliptic quadric) (see, e.\,g., \cite{hir}). The ten three-qubit Pauli matrices of the pentagram are thus represented by ten lines joining the five points of the ovoid in pairs --- as illustrated in Figure 4.

\begin{figure}[t]
\centerline{\includegraphics[width=8.cm,clip=]{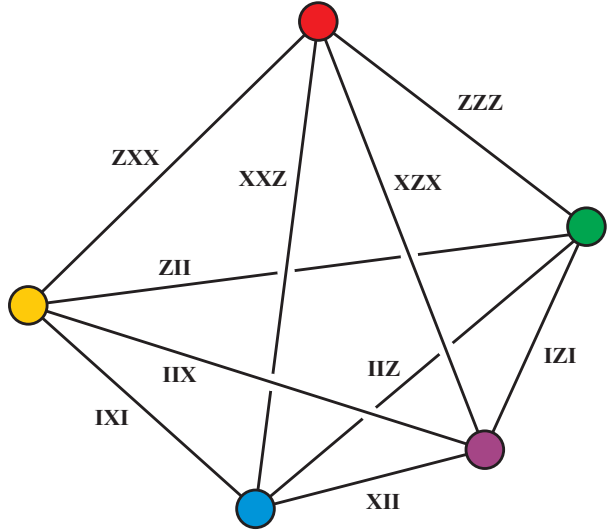}}
\vspace*{.2cm}
\caption{The PG$(3,\,2)$-image of the Mermin pentagram under the Klein correspondence. The edges of the pentagram correspond to the points of an ovoid/elliptic quadric, and its ten observables map into the lines joining the points in pairs.}
\end{figure}

We started this section by the observation that the four observables sharing an edge of the pentagram do not represent a maximal commuting set in the corresponding three-qubit Pauli group. We also stressed that any such maximal set consists of seven elements carried by a Fano plane of $W(5,2)$.  It is, therefore, instructive to see explicitly the projective closures of the edges (up to signs):
\begin{eqnarray*}
&\{ZZI, ZIZ, IZZ\} &({\rm for~green~edge}),  \\
&\{XXI, XIZ, IXZ\} &({\rm for~blue~edge}),  \\
&\{XIX, IZX, XZI\} &({\rm for~violet~edge}),  \\
&\{IXX, ZXI, ZIX\} &({\rm for~yellow~edge}),  \\
&\{IYY, YIY, YYI\} &({\rm for~red~edge}),
\end{eqnarray*}
where $Y \equiv i\sigma_{y}$. One can readily check that, up to a sign, the product of each of the five triples of observables is the identity matrix and so each of them represents indeed a line in the associated Fano plane (which is highlighted in Figure 2, left, by three big bullets).
One further notes that only the observables of the red-edge projective closure contain the matrix $Y$, which reflects the fact that this edge stands on a different footing that the other four.

\section{Conclusion}
We have shown that the Mermin pentagrammatic framework of ten three-qubit observables, which furnishes a very economic and elegant proof of the Kochen-Specker theorem, can be viewed as a distinguished object of finite geometry, namely an ovoid (elliptic) quadric of PG$(3,2)$; the five edges of the pentagram correspond to the five points of the ovoid and its ten vertices/observables are represented by the lines joining pairs of points of the ovoid. This finding may serve as another justification of our firm belief that all important sets of operators/observables associated with finite-dimensional Hilbert spaces are underlaid by notable objects of finite geometry. In the light of this paper, a particularly interesting and challenging task would be to look for higher-rank analogues of Mermin's configuration(s). Already the next case in the hierarchy, $N=4$, deserves serious attention. This is mainly because an associated hyperbolic quadric $Q^{+}(7,2)$, the locus of symmetric elements of the four-qubit Pauli group, is unusual in that it admits a graph automorphism of order
three known as a triality that swaps its points and two systems of generators, and preserves the set of totally singular lines (see, e.\,g., \cite{tits}).

\section*{Acknowledgement}
This work was partially supported by the VEGA grant agency projects 2/0092/09 and 2/0098/10. We are extremely grateful to our friend Petr Pracna for electronic versions of the figures.

\end{document}